\documentclass[twocolumn,pre]{revtex4}
\usepackage{graphicx}
\usepackage{epsf}
\usepackage{times} 
\usepackage{palatino} 
\usepackage{newcent} 
\usepackage{bookman}

\newcommand{\half}{\frac{1}{2}}

\newcommand{\bi}{\begin{itemize}}
\newcommand{\ei}{\end{itemize}}
\newcommand{\be}{\begin{equation}}
\newcommand{\ee}{\end{equation}}
\newcommand{\bea}{\begin{eqnarray}}
\newcommand{\eea}{\end{eqnarray}}
\newcommand{\ba}{\begin{array}}
\newcommand{\ea}{\end{array}}

\newcommand{\npclass}{{\bf NP}}

\begin{document}

\title{Freezing in random graph ferromagnets}
\date{\today}
\author{Pontus Svenson \\
tfkps@fy.chalmers.se}
\affiliation{Institute for Theoretical Physics \\
 Chalmers University of Technology and G\"oteborg University \\
 SE-412 96 Gothenburg, Sweden}

\begin{abstract}
Using $T=0$ Monte Carlo and simulated
annealing simulation, we study the energy relaxation of
ferromagnetic Ising and Potts models on random graphs.
In addition to the expected exponential decay to a zero energy ground
state, a range of connectivities for which
there is power law relaxation and freezing to a metastable state is found. 
For some
connectivities this freezing persists even using simulated
annealing to find the ground state. The freezing is caused
by dynamic frustration in the graphs, and is a feature of the
local search-nature of the Monte Carlo dynamics used.
The implications of
the freezing on agent-based complex systems models are briefly considered.
\end{abstract}

\maketitle

The way a physical system approaches equilibrium is a subject
of interest to both physicists and mathematicians.
In order to measure thermodynamical properties of systems it
is important to be certain that the system really is in
equilibrium. To ensure this, in computer simulations using the
Monte Carlo dynamics it is necessary to first
run the simulation for a long time before measuring.
The way that various properties (e.g., the energy) of the system change
during equilibration
is also interesting in itself, e.g., in studies of how an epidemic
disease or an opinion spreads in a model of social agents.

Here we study the relaxation of the
energy of ferromagnetic Ising and Potts models on random
graphs using Monte Carlo simulations with the 
Metropolis dynamics. We find
an interesting transition as the connectivity
of the graph is varied. For very small connectivities,
the energy relaxes exponentially fast, for intermediate
connectivities the system freezes in a local minimum (with power law
relaxation to it), and for graphs with large connectivities
there is again exponentially fast decay.

The model under consideration is the standard Ising model with
ferromagnetic interactions but with spins placed on
a random graph. In graph theory terminology~\cite{bollobas},
the ensemble used is ${\cal G}(N,M)$, which consists
of all graphs with $N$ vertices
and $M=\half \gamma N$ randomly selected edges. On average, each node is
connected to $\gamma$ others; $\gamma$ is the
{ connectivity} or { average degree} of the graph.
Each edge in the graph is a ferromagnetic interaction between the
two linked spins, and the energy of the model can be taken to be
\be
\epsilon = \frac{1}{N} \sum_{i< j} J_{ij} (1 - \delta_{s_i}^{s_j})
= - \frac{1}{2N} \sum_{i<j} J_{ij} s_i s_i  + \frac{1}{4} \gamma,
\label{hamiltondef}
\ee
where exactly $M$ of the $J_{ij}$'s are non-zero and equal to 1.
Thus, $\epsilon$ counts the number of edges linking 
spins with different values. Note that this differs from the
standard ferromagnetic Hamiltonian by a $\gamma$-dependent term.
Similar models on random graphs have been used to study
many different systems in biology and social science as well as in 
physics (e.g.,~\cite{jainkrishna,li:lectures}).

The model can also be viewed as a constraint satisfaction
problem. Each of the $M$ edges in the graph corresponds
to a constraint that the two linked spins should be equal.
A natural interpretation of this is a model of a
social system where there are $N$ agents choosing from two
different opinions or activities. A link between two agents
would mean that the two prefer to agree. 

By relaxation of a model, we mean the behaviour of the energy
after a quench from a high temperature disordered spin configuration.
The Monte  Carlo method~\cite{metropolis}
tries to decrease the energy of the system by changing the
configuration of spins locally. In the Glauber 
dynamics~\cite{glauber:dynamics} used
in this paper, the change is accomplished
by attempting to flip a randomly selected spin.
If the new spin configuration has lower energy than the old, 
it is accepted. If the
energy is raised $\Delta$ units by the change, the new configuration
is accepted with probability $\exp[-\beta \Delta]$, where
$\beta=1/T$ is inverse temperature (this is the Metropolis~\cite{metropolis}
algorithm). In temperature $T=0$ simulation,
no changes that raise the energy are accepted. In most of the simulations
reported here, the Mitchell-Moore additive generator (see, e.g.,~\cite{knuth2})
was used to generate random numbers. Some runs were also performed using
the standard C library's {\tt drand48()} generator; these gave
the same results.

For the standard 2D Ising model, with $J_{ij} = 1$ if and only if
spins $i$ and $j$ are nearest neighbours on the square lattice, two
behaviours of the relaxation are possible. If the order parameter is
conserved by the dynamics, so that the magnetisation of the system
does not change, $\epsilon \sim t^{-1/3}$, while 
$\epsilon \sim t^{-1/2}$ if single spin flip dynamics are used.
These behaviours can be understood by considering domains of
up and down spins~\cite{bray:coarsening}.

Since a random graph is locally tree-like, it is natural to
approximate the behaviour of the random graph model with
that of the same model on a tree. Johnston and  
Plech\'{a}\u{c}~\cite{johnstonplechac:loops} have
shown that the 
thermodynamical behaviour of the ferromagnetic Ising model
is independent of the presence of loops in a graph: it is the
same on a random regular graph and on a Bethe tree with the same connectivity.
Da Silva and Silva~\cite{dasilvasilva}
studied relaxation in the Ising ferromagnet on { Cayley trees},
and showed that mean field theory predicts exponential relaxation.
Exponential relaxation can also be argued for easily by writing a
mean-field
equation for the time dependence of a spin in terms of its nearest
neighbours.
Glassiness in the Cayley tree ferromagnet has been
studied by Melin~et~al~\cite{melinetal}, who find a crossover
temperature that scales inversely with the logarithm of the
number of surface sites. For the random graph model considered here,
this is 0, since there are no surface sites.

\begin{figure}
\centering
\leavevmode
\includegraphics[width=.75 \columnwidth]{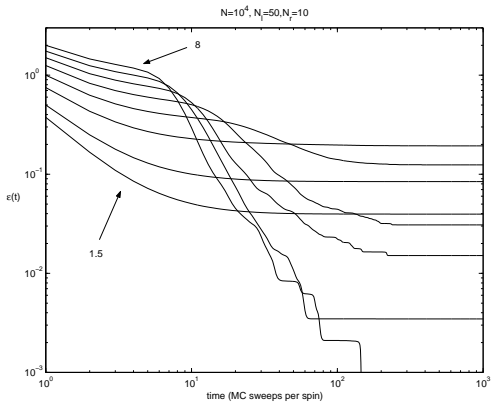}
\vspace*{2mm}
\caption{The relaxation in a ferromagnetic
Ising model on a random graph
with $10^4$ vertices, averaged over 50 graphs and 10 restarts per graph.
For small (not shown) and large $\gamma$, there is a
fast exponential relaxation, while the behavior for $\gamma=2$ and
3 is a power law $\epsilon = \epsilon_0 + t^{-\nu}$, with $\nu \approx 1.3$.
The lower arrow indicates the curve for $\gamma=1.5$ and the upper
one the $\gamma=8$ data. In between them are data for $\gamma=2$, 3, 4,
5, 6, and 7.}
\label{ferrocolenergyfigur}
\end{figure}

Figure~\ref{ferrocolenergyfigur} shows the relaxation behaviour of
$\epsilon$ for $\gamma=1.5$ to 8 and graphs of size $10^4$. 
All data were averaged over 50 different graphs, and the MC algorithm
was restarted in 10 different initial spin configurations for each graph.
In order to check self-averaging, we also made runs
with averages over 5 graphs and 100 initial configurations, and
500 graphs and 1 initial configuration, and found no differences. 
Error bars were determined to be on the order of $10^{-3}$ or smaller. 
The figure shows that large $\gamma$'s cause
very fast relaxation to the ground state, while 
the system freezes for intermediate values of $\gamma$ .
For very small $\gamma$'s,
the relaxation is of course still fast (not shown in the figure).
The behaviour for intermediate values of $\gamma$ is thus different from the
tree-like models. 
We have also obtained similar results using ferromagnetic
$k$ state Potts models on random graphs.

\begin{figure}
\centering
\leavevmode
\includegraphics[width=.5 \columnwidth]{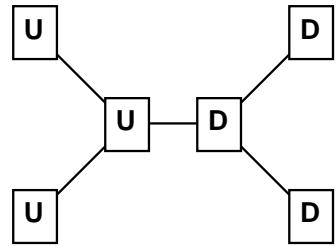}
\vspace*{3mm}
\caption{An example of a configuration that can cause freezing due
to insufficient clustering. U and D denote up and down
spins, respectively.}
\label{frysningsgraf}
\end{figure}

This behavior can be explained qualitatively by
noting that even though the ferromagnetic models always have a
ground state with zero energy,
it is possible for the $T=0$ Monte Carlo algorithm (and all other
local search methods)
to get stuck in a local minimum. The simplest case where this
can happen is when there is a link between two nodes that have different
values and each of the nodes have two neighbours with
the same value, see figure~\ref{frysningsgraf}.
Because there is only one path between the up and down
domains in this figure, it is not possible to lower
the energy by flipping a spin.
Thus, even though the model itself is
solvable and contains no frustration, the dynamics gives
rise to {dynamical frustration} for local search methods.
(Very recently,
Spirin~et~al~\cite{spirin} have found freezing to a blinker
state in the 3-dimensional Ising model. Blinkers will appear
in the random graphs studied here too, but because of the
relatively small connectivities at which the freezing appears 
it is more likely that it is caused by subgraphs such as those
shown in figure~\ref{frysningsgraf}.)

If there are sufficiently many edges between the up and down domains, 
the relaxation will be fast.
No dynamical frustration will occur and one dominant value will
spread quickly through the graph. If there are few edges,
this will take longer, and different
values will dominate different parts of the graph.
This makes it plausible that introducing a metric
and adding a restriction to the range of
the edges could
cause changes in the relaxation.
This conjecture is confirmed by simulations of a model
where the spins are arranged on a chain and edges only allowed
between spins whose distance is less than $\alpha N$, where $\alpha$ is
independent of $N$.
The large
$\gamma$ behavior is now power law relaxation and freezing.
This is similar to the behaviour of the antiferromagnetic 
Ising and Potts models on a random graph~\cite{nprelax}. This is a model
for the graph colouring problem, a combinatorial optimisation 
problem that is \npclass-complete~\cite{papadimitriou},
meaning that its worst case
instances in all likelihood require exponential time to solve
on a deterministic Turing machine.

\begin{figure}
\centering
\leavevmode
\includegraphics[width=.75 \columnwidth]{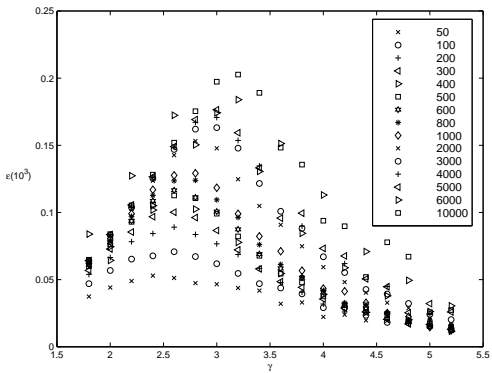}
\caption{This figure shows $\epsilon$ after $10^3$ MC steps per spin
as a function of $\gamma$ for system sizes $N$ ranging from 50 to $10^4$.}
\label{frysenergigamma}
\end{figure}

Returning to the model with no restrictions on the edges,
figure~\ref{frysenergigamma} shows the value of $\epsilon$ after
$10^3$ MC steps per spin as a function of $\gamma$ and
for system sizes ranging from 50 to $10^4$, determined using
about 100 different graphs and about 50 runs for each graph.
The freezing region can be seen clearly.
The error bars of the results shown in this and the other
figures were small, typically on the order of or smaller than
the symbols used to plot the data.

\begin{figure}
\centering
\leavevmode
\includegraphics[width=.75 \columnwidth]{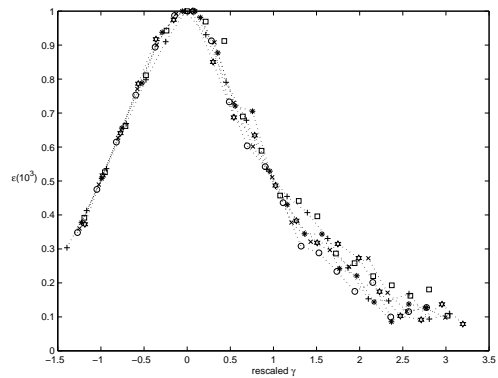}
\caption{Some of the data from figure~\ref{frysenergigamma}, rescaled
by the maximum energy for each $N$ and plotted as a function of
a rescaled parameter $\hat{\gamma}$ described in the text.}
\label{frysmaster}
\end{figure}

For large $N$, it is possible to fit all the data from
figure~\ref{frysenergigamma} on a universal curve.
Figure~\ref{frysmaster} shows the energy for large $N$, rescaled
so that the maximum is 1, as a function of a rescaled parameter
\be
\hat{\gamma} = \frac{\gamma-\gamma_0}{\Delta^{\pm}}
\ee
where $\gamma_0$ is the location of the maximum and $\Delta^{\pm}$
was calculated so that
$\hat{\gamma}=\pm 1$ marks the points where the energy attains
half its maximum value. Note that the original data are non-symmetric
around their maxima: when calculating $\hat{\gamma}$ we divided by
different factors right and left of the maximum. To determine
the locations of the maximum and $\half$-maximum points,
a cubic spline fit of the data was used; the plot shows the original data
points.

Since the energy barrier surrounding the local minimum shown 
in figure~\ref{frysningsgraf} is small, it is likely that
finite temperature MC simulations would not show the same behaviour.
To test this, we have also tried simulated
annealing~\cite{kirkpatrickgelattvecchi} on the problem. Simulated
annealing starts at a high temperature and then gradually
decreases it during the simulation. We used a linear
decrease in temperature, $T(t) = T_1 - k t$, where $k$ was
chosen so that the simulation ends at zero temperature.

\begin{figure}
\centering
\leavevmode
\includegraphics[width=.75 \columnwidth]{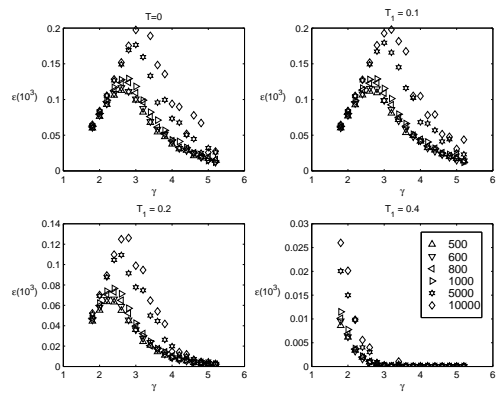}
\caption{This figure shows the energy after $10^3$ MC steps per spin
for system sizes from 500 to $10^4$ using $T=0$ MC simulation (top left)
and simulated annealing with start temperature $T_1=0.1$, 0.2, and 0.4
(top right, bottom left, bottom right) and end temperature 0.
Most of the freezing effect disappears using simulated annealing,
but even for the $T_1=0.4$ runs the algorithm was unable to find
the ground state for $\gamma \sim 2$. }
\label{annealingfrysenergigamma}
\end{figure}

Figure~\ref{annealingfrysenergigamma} compares the values
of $\epsilon$ after $10^3$ MC steps per spin for the $T=0$
MC algorithm with those obtained using simulated annealing with
start temperature $T_1=0.1$, $0.2$, and $0.4$ and
the same averaging as in the previous runs.
Most of the freezing disappears in these runs, but there is a region of
$\gamma$'s for which it remains. Note that the percolation threshold
for random graphs is at $\gamma = 1$, well below the freezing.

In conclusion, we presented results from
Monte Carlo and simulated annealing studies of the ferromagnetic
Ising model on random graphs. We find different regions of
behaviour of $\epsilon(t)$ --- the expected exponential relaxation
but also some regions where there is power law relaxation.
More importantly, freezing was found in the model. The freezing
persisted even for some simulated annealing runs, but almost
disappears for large start-temperatures. The freezing is a feature of
the local search and hill-climbing characteristics of the MC method 
used. This has implications
for the study of models of choice-making agents on random graphs:
for some connectivities it is not possible to reach a consensus
or the globally most effective solution by using only local 
information. There are
intriguing similarities and differences between this model
and the corresponding antiferromagnetic model studied
elsewhere; these could be studied further by examining the
model where there are mostly ferromagnetic bonds but with some
probability $p$ of instead having an antiferromagnetic bond.

Acknowledgements: I thank Stellan \"Ostlund for commenting
on a previous version of this manuscript.

\end{document}